# Single-shot pump-probe technique by the combination of an echelon and a grating with a time window of 109 ps

Tianchen Yu [a], Junyi Yang [a,*], Zhongguo Li [b], Xingzhi Wu [c], Yu Fang [c], Yong Yang [d,*], Yinglin Song [a,*]

[a]School of Physical Science and Technology, Soochow University, Suzhou 215006, China

[b]School of Electronic and Information Engineering, Changshu Institute of Technology, Changshu 215500, China

[c]Jiangsu Key Laboratory of Micro and Nano Heat Fluid Flow Technology and Energy Application, School of Physical Science and Technology, Suzhou University of Science and Technology, Suzhou 215009, China

[d]School of Optoelectronic Science and Engineering, Soochow University, Suzhou 215006, China

**\*Corresponding author**
E-mail address: yjy2010@suda.edu.cn (Junyi Yang), yangy@suda.edu.cn (Yong Yang) ylsong@suda.edu.cn (Yinglin Song)

**Abstract**

In this study, using only a single pulse, pump-probe measurement with a large time window of more than 100 ps is implemented. A commercial grating is used to encode a time window of ~ 56 ps in a single pulse; therefore, there is no need for machining customization. In addition, in this technique, the grating surface is accurately imaged, eliminating the image blur problem caused by phase differences in previous echelon-based techniques. Moreover, to make full use of the grating surface and obtain a larger time window, a simple reflection echelon is combined that matches the grating in the time window. This combination encoding strategy results in a total time window of ~ 109 ps and maintains accurate imaging of the grating surface. This time window is an order of magnitude greater than the maximum reported values of the echelon encoding strategy and the angle beam encoding strategy. To demonstrate this single-shot pump-probe technique, the two-photon absorption process of ZnSe and the excited-state absorption process of a symmetrical phenoxazinium bromine salt were studied. The possibility of further improving the experimental setup is also discussed.

## 1. Introduction

The pump-probe technique is a powerful tool for diagnosing ultrafast dynamics in all kinds of materials. Typically, the pump-probe technique uses a strong beam (pump beam) to excite the material, followed by a weak beam (probe beam) to diagnose the material. In the measurement, the optical path difference (time delay) between the two beams is adjusted consecutively by a mechanical time delay stage. Finally, a curve of the probe light transmittance to the time delay is obtained, i.e., the dynamic trace of the material. However, this technique is not suitable for the diagnosis of many irreversible ultrafast phenomena, such as photodissociation [1, 2], phase transition [3], laser ablation [4] and explosion [5], due to the repeated excitation of multiple pump pulses at each time delay. In addition, for a long-period measurement, fluctuations in the energy from pulse to pulse can increase the noise in the entire dynamic trace. Therefore, to obtain an adequate signal-to-noise ratio, thousands of pulses must be averaged for each time delay.

A variety of single-shot pump-probe techniques have been proposed. In contrast to the traditional pump-probe technique, the single-shot pump-probe technique requires only a single pump pulse to excite the material followed by a single probe pulse to record the entire dynamic trace of the material. In general, the single-shot pump-probe technique encodes the time delay into a single probe pulse using spatial or spectral strategies. Currently, the commonly used encoding strategies include the angle beam encoding strategy [1, 2, 6-8], spectral encoding strategy [9-12], and echelon encoding strategy [3, 5, 13-16]. The angle beam encoding strategy utilizes the crossed line focus of the pump beam and probe beam on the sample plane, and the time delay is encoded along the focal line. Using this strategy, the maximum time window reaches 60 ps [8]. However, a large sample area with high homogeneity is required for the angled beam encoding strategy. For the spectral encoding strategy, the sample dynamic information is recorded by the different spectral components of the chirped probe pulse. The spectral encoding strategy can implement a high acquisition rate because no camera is required for imaging, and the time window depends on the width of the chirped pulse. Therefore, the time window can reach more than 100 ps [12] using this strategy. However, the minimum time resolution $\tau_R = \sqrt{\tau_0 \tau}$ depends on the transform-limited pulse $\tau_0$ and the chirped pulse $\tau$, so there is a trade-off between a large time window and a short time resolution [10]. In addition, an oscillation pattern near the origin of the spectral encoded dynamic trace may distort the actual dynamic trace [12]. For the echelon encoding strategy, the probe pulse is encoded as a subpulse chain by the reflection or transmission echelon. The step difference and the echelon height (total step difference) determine the minimum time resolution and the time window, respectively. This strategy results in a well-calibrated time delay. However, image blurring problems can occur when a large time window is needed. This is due to the large time window requiring a high echelon height. For example, a 100 ps time window requires the echelon height to be at least 1.5 cm for a reflection echelon. This results in large phase differences and makes it difficult to accurately image all the step surfaces simultaneously on an image plane. Therefore, the maximum time window reported by the echelon encoding strategy is ~ 37.7 ps [15], which is shorter than that of the other two strategies.

In addition to the above encoding strategies, the use of a grating or a digital micromirror device (DMD) has shown promise [17-20]. However, few studies on this topic exist, and the time window, to the best of our knowledge, has not exceeded 30 ps. In this work, therefore, we propose a single-shot pump-probe technique with a large time window using a combination encoding strategy. A combination of a simple reflection echelon and a commercial grating is used to spatially encode the single probe pulse, and a time window of ~ 109 ps is obtained while maintaining accurate imaging of the grating surface. This time window is an order of magnitude greater than the maximum reported values of the angle beam encoding strategy and the echelon encoding strategy. To demonstrate this technique, we studied the ultrafast two-photon absorption process of ZnSe and the long-life excited-state absorption process of a symmetrical phenoxazinium bromine salt.

## 2. Experimental setup

A schematic of the single-shot pump-probe technique is shown in Fig. 1. A mode-locked Yb: KGW-based fiber laser (1030 nm, 190 fs full width at half maximum (FWHM), 20 Hz) was used as the laser source. The laser output wavelength is frequency doubled to 515 nm in this study. The laser beam is divided into a strong pump beam and a weak probe beam using a beam splitter ($BS_1$). The optical path of the pump beam can be adjusted by a time delay stage. It should be noted that the time delay stage is used only to compensate for the initial optical path difference between the pump and probe beams, and the position of the time delay stage is fixed in each measurement, which is significantly different from the role of the time delay stage in traditional pump-probe measurements. The pump beam is focused on the sample by lens $L_9$ ($f$ = 500 mm). The energy of the pump and probe beam is adjusted by the neutral density attenuator (ND). The probe beam is expanded to a beam spot approximately 50 mm in diameter by lenses $L_1$-$L_4$. Apertures $A_1$ and $A_2$ intercepted the central part of the beams.

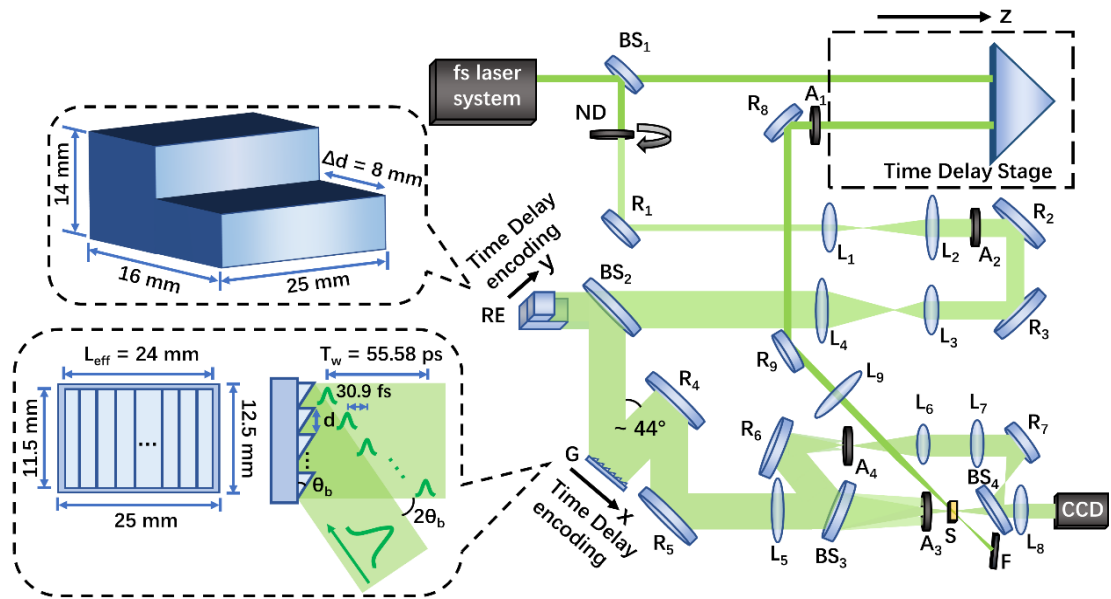
4

**Fig. 1.** Schematic of the single-shot pump-probe technique. $BS_1$-$BS_4$: beam splitter; $R_1$-$R_9$: mirror; $L_1$-$L_9$: lens; RE: reflection echelon; G: grating. $A_1$-$A_4$: apertures; ND: neutral density attenuator; S: sample plane (focal plane); F: flapper; CCD: charge coupled device; top dotted box: schematic of the RE; Δd: step difference; lower dotted box: grating size and spatial encoding; $\theta_b$: blaze angle; $L_{eff}$: effective length of the grating; d: groove width; $T_w$: time window.

A reflection echelon (RE) spatially encodes the single probe pulse as two subpulses in the step direction (y-direction). As shown in the upper dotted box, the RE consists of two steps with a step difference Δd of 8 mm. Therefore, at a normal incidence, the RE can encode a time delay of 2Δd/c = 53.33 ps between the two subpulses in the y-direction, where c is the light speed. The encoded probe beam then illuminates a ruled grating (G) by a beam splitter ($BS_2$). The groove direction of the grating is perpendicular to the step direction of the echelon. As shown in the lower dotted box, the grating size is 12.5 × 25 mm² with a ruled area of 11.5 × 24 mm² and 75 grooves per mm. At an incident angle approximately twice the blaze angle $\theta_b$ (22.02°), the direction of the geometric reflection beam is parallel to the grating normal, and a set of time delays is encoded along the grating length (x-direction).

Therefore, the ruled grating further spatially encodes the two parallel subpulses in the x-direction. The encoded time window $T_w$ of each subpulse can be calculated by:

$$T_w = L_{eff} \sin(2\theta_b)/c \qquad (2\text{-}1)$$

Here, $L_{eff}$ is the effective length of the grating. This strategy is somewhat similar to the angle beam encoding strategy. However, the time delay is encoded on the grating plane in our geometry instead of on the sample surface. From Eq. (2-1), to obtain a large time window, we need to increase the effective length $L_{eff}$ and choose a large blaze angle $\theta_b$. As shown in Fig. 2(a), the calculated time window of the grating increases with $L_{eff}$ and $\theta_b$. In our experiment, the time window $T_w$ is calculated as ~ 55.58 ps (indicated by the asterisk) with $L_{eff}$ = 24 mm and $\theta_b$ = 22.02°. However, for the ruled grating, we need to consider the ratio $r_s$ of the illumination area to the groove area. In our configuration, the ratio $r_s$ can be calculated by:

$$r_s = 1 - 2(\sin\theta_b)^2 \qquad (2\text{-}2)$$

As shown in Fig. 2(b), a larger $\theta_b$ will result in a lower illumination ratio $r_s$. In our experiment, $r_s$ is ~ 72% (indicated by the asterisk). Therefore, a ruled grating with an effective length $L_{eff}$ of 24 mm and a blaze angle $\theta_b$ of 22.02° is suitable for obtaining both a large time window $T_w$ and a high illumination ratio $r_s$. The encoded time interval $\Delta t$ can also be calculated by $\Delta t = d\sin(2\theta_b)/c$ with groove width $d$. In our experiment, the time interval $\Delta t$ is ~ 30.9 fs which is shorter than the pulse width of 190 fs. Therefore, this time interval is not the minimum time resolution in our experiment.

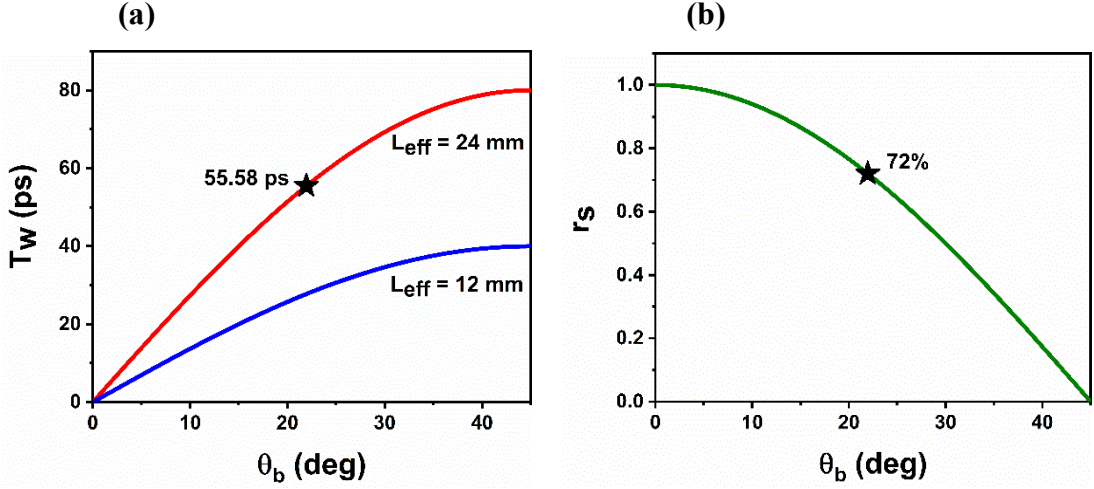

**Fig. 2.** (a) Calculated time window $T_w$ of the grating with the effective length $L_{eff}$ and the blaze angle $\theta_b$; (b) illumination ratio $r_s$ of the groove with the blaze angle $\theta_b$.

Overall, with the combination of the reflection echelon RE and the ruled grating G, a two-dimensional spatial encoding with a total time window of ~ 109 ps is generated. In addition, there is a time overlap of ~ 2.25 ps between the two subpulses of the RE, considering that the time window for each subpulse is ~ 55.58 ps and the time delay is ~ 53.33 ps. The encoded probe beam is focused by lens $L_5$ ($f = 250$ mm) to interrogate the pumped area and then imaged on a CCD (spinnaker, 16 bit, 2448×2048 pixels, 3.45×3.45 μm² each pixel) using lens $L_8$ ($f = 80$ mm). The exposure time of the CCD is set to 50,000 μs for single-shot acquisition. In addition, a reference beam is divided using a beam splitter (BS$_3$) and imaged on the CCD reduced by a factor of 2 by lenses $L_6$ ($f = 50$ mm) and $L_7$ ($f = 100$ mm), and the reference image is inverted. The reference image is used to monitor the spatial fluctuation of the pulse energy. In our configuration, the grating surface is accurately imaged on the CCD. Therefore, dynamic information can be extracted while avoiding the image blur problem by phase differences in the traditional echelon encoding strategy since spatial encoding is completed at the grating surface. Moreover, to account for the diffraction of grating G, filtering aperture A$_3$ is placed in front of sample plane S (focal plane) at a distance of 6 cm. Similarly, a filtering aperture A$_4$ is placed on the focal plane of the reference path. The diffraction light of the grating follows the following equation:

$$m\lambda = d(\sin\alpha + \sin\beta) \qquad (2\text{-}3)$$

where $\lambda$ is the wavelength, $m$ is the diffraction order, and $\alpha$ and $\beta$ represent the incidence and diffraction angles, respectively. When $\beta = 0$, the separation between the diffraction orders is $\Delta L = \lambda f / d$, where $f$ is the focal length of lens $L_5$. In our experiment, $\Delta L$ is ~ 0.97 cm, and the probe spot size is ~ 5.76 mm in the *x*-direction at a distance of 6 cm from the focal plane. Therefore, the filtering aperture $A_3$ with an aperture diameter of 1.10 cm is used to filter the diffractions.

## 3. Experimental results

### 3.1 Two-photon absorption (TPA) measurements

Polycrystalline ZnSe with a thickness of 3 mm was used to study the TPA process. In the experiment, the sample position was offset from the focus of lens $L_9$ to obtain a pump spot of ~ 2 mm, which completely covered the probe spot (~ 0.8 mm in the *x*-direction and ~ 23 μm in the *y*-direction). The pump and probe energies are 10 μJ and 3 nJ, respectively. In each measurement, four sets of images were acquired: i) the grating image when both the pump and probe beams were turned on, i.e., the "pump on" image; ii) the grating image when the pump beam was turned off while the probe beam was turned on, i.e., the "pump off" image; iii) the environmental background when both the pump and probe beams were turned off; and iv) the pump beam background when the pump beam was turned on while the probe beam was turned off.

    The TPA process derives from the bound electron response to an external photoelectric field and typically occurs within a few femtoseconds. Therefore, the dynamic trace of the TPA represents the cross-correlation between the pump pulse and the probe pulse in our experiment and appears only near the position of the zero-time delay (i.e., the moment when the pump and probe beams meet). If we change the *z*-position of the time delay stage (as indicated by the arrow in Fig. 1) after each measurement, the position of the zero-time delay will change correspondingly, and this can help us calibrate the time window of the grating. As shown in Fig. 3, a set of grating images was obtained after several sets of measurements. In Fig. 3(a), a clear grating image of the "pump off" is shown and the environmental background was subtracted. The upper part is the reference image, and the lower part is the signal image. As mentioned in Section 2, the reference image is inverted relative to the signal image and

reduced by a factor of 2. The grating is divided into two rows by the reflection echelon with a narrow diffraction boundary. Therefore, the first (upper) and second (lower) rows have different time delays. In addition, the grating intensity distribution is not completely uniform. This does not affect the measurement results, as the probe beam is much weaker than the pump beam, and the intensity variation is normalized and corrected during the subsequent data processing. The corresponding "pump on" image is shown in Fig. 3(b). Similarly, we subtracted the pump beam background. A strip region appears (indicated by the red arrow) in the "pump on" image, which is due to the TPA process occurring near the zero-time delay.

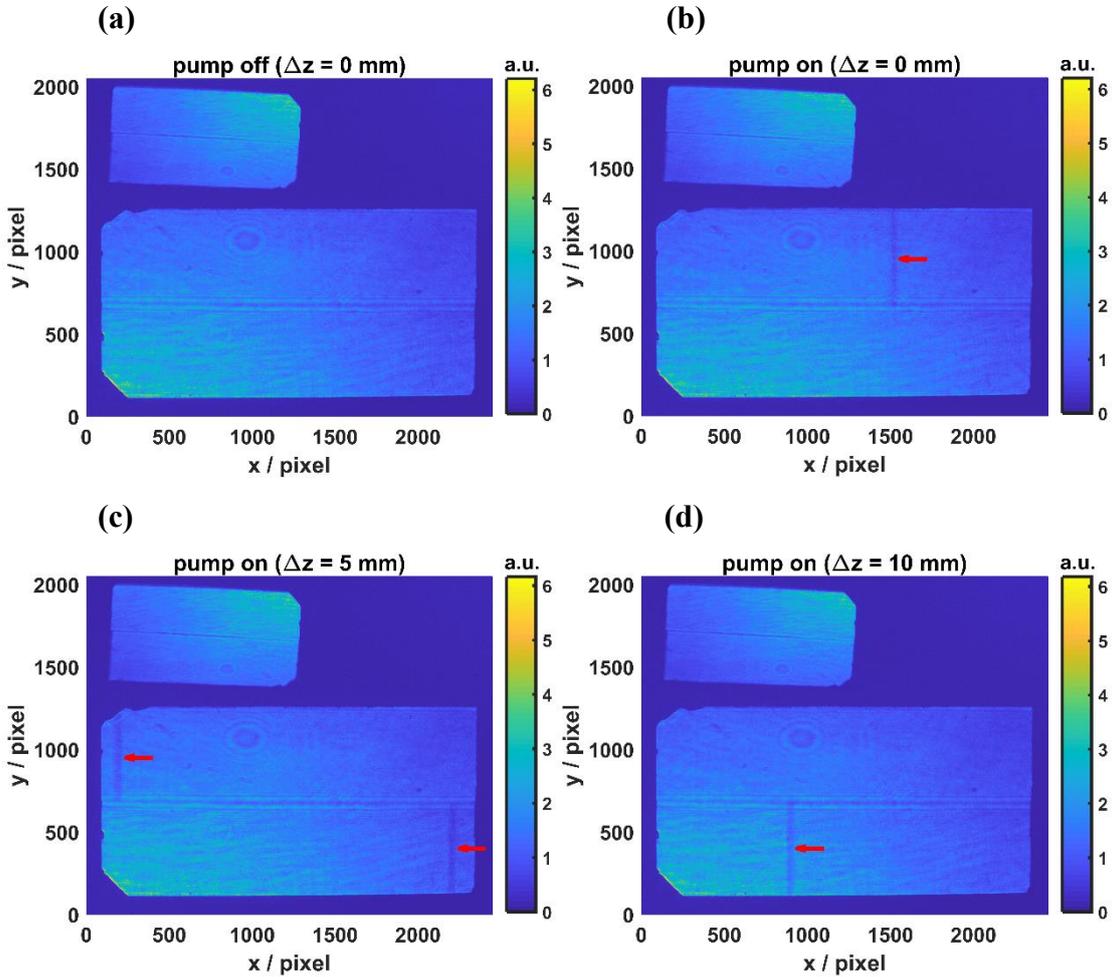

**Fig. 3.** (a) "Pump off" image at $\Delta z = 0$ mm; (b) "pump on" image at $\Delta z = 0$ mm; (c) "pump on" image at $\Delta z = 5$ mm; (d) "pump on" image at $\Delta z = 10$ mm. The red arrow indicates the movement direction of the TPA or the zero-time delay.

We set the corresponding z-position of the time delay stage as $\Delta z = 0$ mm and changed it in the next measurement. The "pump on" images at $\Delta z = 5$ mm and 10 mm were obtained as shown in Fig. 3(c) and 3(d), corresponding to time delay increases of 33.3 ps and 66.7 ps, respectively. The movement of the TPA indicates that the time delay of the grating increases from the right to the left and from the first row to the second row, as indicated by the direction of the red arrow. In addition, the TPA appears in both the first and second rows at Δz = 5 mm. This is as we expected due to the overlap in the time windows of the two rows. To extract the dynamic traces of ZnSe, the first and second rows were processed separately as follows:

$$T_{nor}^{S} = \frac{S_{11} - S_{10}}{S_{01} - S_{00}} \quad (3\text{-}1)$$

$$T_{nor}^{R} = \frac{R_{11} - R_{10}}{R_{01} - R_{00}} \quad (3\text{-}2)$$

$$T_{cor} = \frac{T_{nor}^{S}}{T_{nor}^{R}} \quad (3\text{-}3)$$

Here, $S$ and $R$ represent the signal image and reference image, respectively. For the first subscript, "1" means "pump on", and "0" means "pump off". Similarly, for the second subscript, "1" means "probe on", and "0" means "probe off". $T_{nor}^{S}$ and $T_{nor}^{R}$ represent the normalized transmittance of the signal image and reference image, respectively. It should be noted that the raw data amount of $T_{nor}^{R}$ is only half of the amount of $T_{nor}^{S}$ since the reference image is reduced by a factor of 2. Therefore, we integrate the pixels along the $y$-axis and then perform linear interpolation for $T_{nor}^{R}$. The corrected signal trace $T_{cor}$ is obtained by dividing $T_{nor}^{S}$ by $T_{nor}^{R}$. In Fig. 4, we present the extracted traces of Fig. 3(a) and 3(b), where Δz = 0 mm. The left and right graphs are the traces of the first and second rows, respectively.

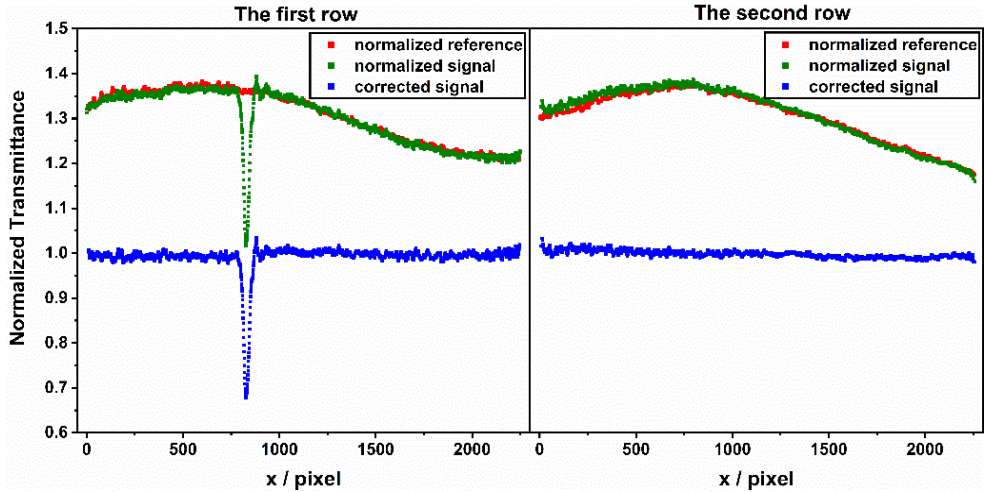

**Fig. 4.** Extracted traces of ZnSe at Δz = 0 mm from the first and second rows. Green square: normalized signal trace; red square: normalized reference trace; blue square: corrected signal trace.

To make this easier to observe, we offset the *y* coordinates of the curve appropriately. Taking the first row as an example, the normalized signal trace $T_{nor}^{S}$ (green square) has a significant TPA valley. However, there is a large fluctuation in the baseline even after normalization. This is due to the spatial fluctuations between the "pump on" and "pump off" images since we acquired them by two independent pulses. Therefore, to correct the spatial fluctuations in pulse energy, a reference image is necessary. As expected, the normalized reference trace $T_{nor}^{R}$ (red square) conforms well to the normalized signal trace $T_{nor}^{S}$ at baseline. Therefore, dividing $T_{nor}^{S}$ by $T_{nor}^{R}$ eliminates the spatial fluctuations from pulse to pulse and yields the corrected signal trace $T_{cor}$ (blue square). In addition, the first and second rows exhibit similar spatial fluctuations. This is because we integrate the pixels along the *y*-axis and retain only the spatial fluctuations along the grating length (*x*-axis). The same processing steps were implemented for Fig. 3c and Fig. 3d. By comparing the *x*-positions of the TPA valley, we determined that the time resolution per unit pixel in both the first and second rows is ~ 25.3 fs/pixel along the *x*-axis. Therefore, the time window of each row is ~ 57 ps and the total time window is ~ 109 ps. This result is consistent with the values calculated in Section 2. In Fig. 5, we present the complete single-shot dynamic traces of ZnSe at Δz = 0 mm and 10 mm after calibrating the time window. Both traces exhibit an ultrafast TPA process within the time window of 109 ps, and from Δz = 0 mm to 10 mm, the TPA valley shifts by ~ 66.7 ps.

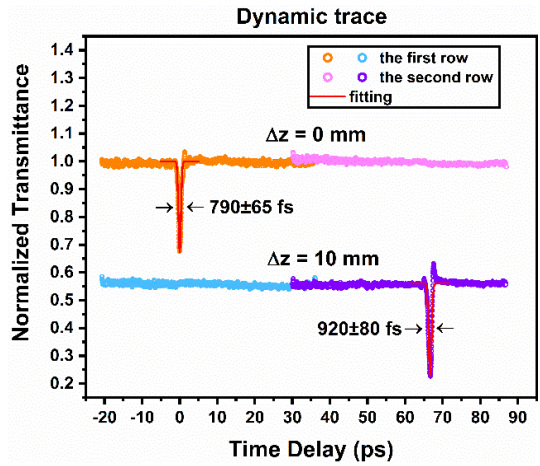

**Fig. 5.** Complete single-shot dynamic traces of ZnSe at Δz = 0 mm and 10 mm. Orange and blue hollow circles: extracted traces at Δz = 0 mm and 10 mm from the first row, respectively; pink and purple hollow circles: extracted traces at Δz = 0 mm and 10 mm from the second row, respectively; red solid lines: theoretical fitting.

We can use a simple Gaussian function to fit the TPA trace of ZnSe:

$$T_{normalized} = 1 + A_0 \exp\left(-\left(\frac{t}{\tau_{TPA}/2\sqrt{\ln 2}}\right)^2\right) \quad (3\text{-}4)$$

where $A_0$ is a constant determined by the amplitude of the TPA and $\tau_{TPA}$ is the FWHM of the TPA valley. The fitting curves are shown in Fig. 4(b) with red solid lines. The fitting parameters $A_0$ and $\tau_{TPA}$ are -0.325 and 790±65 fs and -0.330 and 920±80 fs for the first and second rows, respectively. The average $\tau_{TPA}$ is ~ 855 fs, which is approximately 3 times the pulse cross-correlation width (269 fs, $\sqrt{2}$ times 190 fs (FWHM)). $\tau_{TPA}$ also represents the minimum time resolution in our experimental setup. Several factors can cause a broadening of the laser pulse or extend the time resolution, including the angular spectral dispersion, pump and probe beam crossing on the sample surface, and sample thickness [7, 19, 20]. To obtain a shorter time resolution, we need to increase the numerical aperture $NA$ of the converging lens L$_5$, reduce the angle $\theta$ between the pump and probe beams, and use a thinner sample (see Fig. S1 and a quantitative calculation in the Supporting Information).

**3.2 Excited-state absorption (ESA) measurements**

Under the same experimental conditions, a symmetrical phenoxazinium bromine salt/ DMF solution [21] with a concentration of 1.60 mg/mL was placed in a quartz cuvette with a thickness of 2 mm to study the ESA process. The ESA process originates from the excited organic molecules transition from the ground state to an excited state by absorbing photons, which further absorb photons in the excited state. To better show the intensity change induced by the ESA in the signal image, Fig. 6 shows the differential signal image ΔI$_{signal}$ of the "pump on" and "pump off" images at a pump pulse energy of 19.58 μJ. A sharp decrease in the light intensity is observed in the first row and this reduction occurs throughout the second row. This indicates that the ESA process of the symmetrical phenoxazinium bromine salt should have a long life of more than 100 ps. Using the same steps as in Section 3.1, we separately extracted the dynamic traces from the first row and second row. The complete single-shot dynamic trace of the symmetrical phenoxazinium bromine salt is shown in Fig. 7.

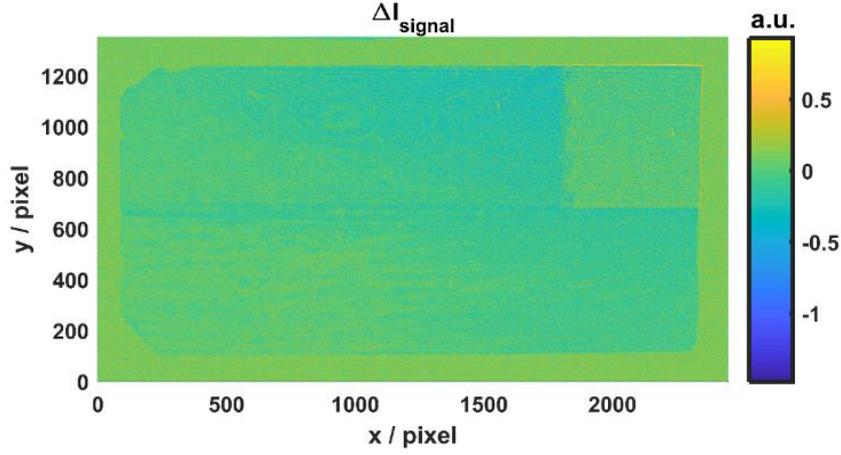

**Fig. 6.** Differential signal image of the "pump on" and "pump off" images at a pump pulse energy of 19.58 μJ.

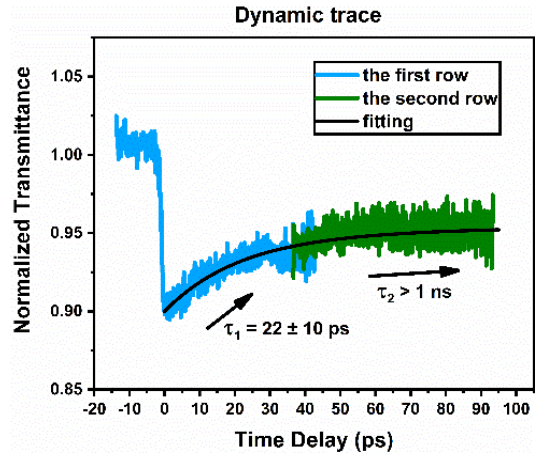

**Fig. 7.** Complete single-shot dynamic trace of the symmetrical phenoxazinium bromine salt/DMF solution. The solid blue line represents the extracted trace from the first row, the solid green line represents the extracted trace from the second row, and the solid black line represents the theoretical fitting.

The dynamic trace of the first row (blue solid line) exhibits a fast recovery process, whereas that of the second row (green solid line) becomes slower. Therefore, the complete dynamic trace can be fitted using a double-exponential function:

$$T_{normalized} = 1 + A_1 \exp\left(-\frac{t}{\tau_1}\right) + A_2 \exp\left(-\frac{t}{\tau_2}\right) \qquad (3\text{-}5)$$

where $A_1$, $\tau_1$ and $A_2$, $\tau_2$ are the amplitudes and lifetimes of the fast and slow processes, respectively. The fitting curve is also shown in Fig. 6 with the black solid line. The fitting parameters $A_1$, $\tau_1$ and $A_2$, $\tau_2$ are -0.05, 22±10 ps and -0.05, > 1 ns, respectively. The two fitted life parameters can be attributed to the singlet excited state $S_1$ and the higher singlet excited state $S_2$ in the symmetrical phenoxazinium bromine salt. This result is consistent with the reports in [21] using a transient absorption spectrum and in [22] using a two-dimensional reflection echelon. Therefore, the results validate the effectiveness of our technique for the measurement of long-life dynamic processes. We can define the single-shot signal-to-noise ratio ($SNR$) of the measurement as $SNR = \Delta T / 2\sigma$; here, we take $\Delta T$ as the average normalized transmittance change in the slow process (time delay $t_d > 70$ ps), and $\sigma$ is the standard deviation in this region. Therefore, $SNR$ is ~ 6.4 in our experiment. Moreover, if we set $SNR = 1$, the minimum relative transmittance change that can be detected in a single-shot is ~ 0.75% in our experiment.

## 4. Discussion

To demonstrate this technique, we chose a degenerate wavelength 515 nm pump-probe scheme, however, different pump-probe wavelengths and polarizations can be selected according to the experimental requirements. Similar to the echelon encoding strategy [3, 14] or the angle beam encoding strategy [1, 2, 8], the grating encoding strategy can also implement wavelength-resolved measurements (single-shot transient absorption spectroscopy). This requires the use of a white light probe and the addition of a spectrometer in front of the CCD. The spectrum is decomposed in the *y*-direction, which is perpendicular to the grating time-axis (*x*-direction). In this way, therefore, the time dimension in the *y*-direction should be replaced by the spectral dimension, which limits the available time window to approximately 50-60 ps. In this study, we focused mainly on the report of a single-shot pump-probe technique with a large time window and accurate imaging. In the current configuration, we only accurately image the grating plane on the CCD, not the reflection echelon. Therefore, there is a diffraction pattern similar to the interference fringes between the rows of the grating (as shown in Fig. 3). During the data processing, we simply discarded the data in this region because it occupies only a small part of the grating plane and does not affect the processing results. In our experiment, the distance between the grating and the echelon is ~ 270 mm. This distance is limited only by the interference between the optical elements in the experiment and can be shortened as much as possible. Another method is to add a 4f imaging system between the reflection echelon and the grating to project the echelon image onto the grating surface. Therefore, even though a two-step echelon was used in this study, more steps could be used, such as four steps, which would increase the total time window to more than 200 ps.

## 5. Conclusion

We demonstrated a single-shot pump-probe technique by combining a simple reflection echelon and a commercial grating. A large time window of ~ 109 ps is obtained, and accurate imaging of the grating surface is maintained, which eliminates the image blur problem caused by the phase differences in the echelon encoding strategy. The obtained time window is an order of magnitude greater than the maximum reported values of the angle beam encoding strategy and the echelon encoding strategy. Studies of the TPA process of ZnSe and the ESA process of a symmetrical phenoxazinium bromine salt demonstrate either ultrafast or slow absorption dynamic processes; this technique can provide reliable measurements. The measured average minimum time resolution is ~ 855 fs in our setup, and the minimum relative transmittance change that can be detected in a single-shot is ~ 0.75%. The time resolution is mainly limited by the angular spectral dispersion, beam crossing, and sample thickness. This technique can also be modified to a wavelength-resolved technique or to further increase the time window to more than 200 ps. In summary, using the echelon and grating combination strategy, a single-shot pump-probe technique with a large time window of more than 100 ps can be achieved for diagnosing various dynamic processes ranging from subpicoseconds to hundreds of picoseconds.

**Acknowledgements.**

We gratefully acknowledge the National Natural Science Foundation of China (11704273, 51607119), National Safety Academic Fund (U1630103), and Science and Technology Innovation Team of Guizhou Education Department (Grant No. [2023]094).

**Conflict of interest.**

The authors declare no conflicts of interest.

# Supporting Information

## Single-shot pump-probe technique by the combination of an echelon and a grating with a time window of 109 ps

Tianchen Yu [a], Junyi Yang [a, *], Zhongguo Li [b], Xingzhi Wu [c], Yu Fang [c], Yong Yang [d, *], Yinglin Song [a, *]

[a]School of Physical Science and Technology, Soochow University, Suzhou 215006, China

[b]School of Electronic and Information Engineering, Changshu Institute of Technology, Changshu 215500, China

[c]Jiangsu Key Laboratory of Micro and Nano Heat Fluid Flow Technology and Energy Application, School of Physical Science and Technology, Suzhou University of Science and Technology, Suzhou 215009, China

[d]School of Optoelectronic Science and Engineering, Soochow University, Suzhou 215006, China

**\*Corresponding author**

E-mail address: yjy2010@suda.edu.cn (Junyi Yang), yangy@suda.edu.cn (Yong Yang) ylsong@suda.edu.cn (Yinglin Song)

One of the main factors that can significantly degenerate the time resolution is the angular spectral dispersion of the grating [1]. The angular spectral dispersion $q$ causes a linear chirp of the pulse. For the diffraction angle $\beta=0$, the angular spectral dispersion is $q=d\beta/d\omega=-\sin\alpha/\omega$. Here, $\alpha$ is the incident angle, $\omega=2\pi c/\lambda$ is the center angular frequency of the laser, and $\lambda$ is the center wavelength. It is assumed that the lens and mirrors do not change the pulse width and the pulse broadening occurs only between the grating and the converging lens $L_5$, as shown in Fig. S1(a).

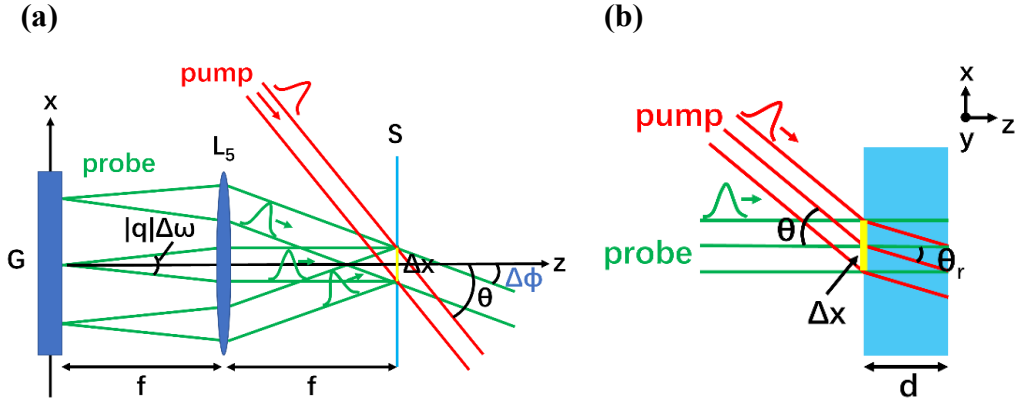

**Fig. S1.** (a) Schematic of the angular spectral dispersion of the grating and beams crossing at the sample plane. (b) Enlarged schematic of the beam crossing in the sample. G: grating; S: sample plane (focal plane); Δx: probe spot on the sample; ΔΦ: angular rotation of the probe beam; θ: cross angle of the pump and probe; $\theta_r$: refraction angle of the pump beam within the sample; d: sample thickness.

The pulse width $\tau$ (half width at 1/e maximum) in the sample plane S (back-focal plane of lens $L_5$) can be expressed as [1]:

$$\tau = \tau_0 \left[1 + (t_c/\tau_0)^4\right]^{1/2} \qquad (1)$$

This formula assumes that the initial pulse has a Fourier transform-limited form $I(t) = I_0 \exp(-t^2/\tau_0^2)$, where $\tau_0$ is the initial pulse width (half width at 1/e maximum). $t_c$ is the characteristic time of the experimental setup:

$$t_c = \left(\frac{|q|T}{2\sin^{-1} NA}\right)^{1/2} \qquad (2)$$

Here, $T$ is the time window for the grating, $NA = \sin(L_{eff}/2f)$ is the numerical aperture of lens $L_5$, $L_{eff}$ is the effective length of the grating, and $f$ is the focal length of lens $L_5$. In our experiments, $\alpha \sim 44°$, $T \sim 56$ ps, $\lambda = 515$ nm, $L_{eff} = 24$ mm, $f$

= 250 mm, and $\tau_0$ is assumed to be 114 fs ($190\text{fs}(\text{FWHM})/2\sqrt{\ln 2}$). Therefore, using Eqs. (1) and (2), we can derive that $t_c$ and $\tau$ are 332 fs and 974 fs, respectively.

The second factor that needs to be considered is the cross angle between the pump and probe beams on the sample surface. As shown in Fig. S1(a), we assume that the pump and probe beams are in the *xz* plane with an angle $\theta$. For convenience, we ignore the small angular rotations $\Delta\Phi$ of the probe beam [1], ~ ±2.7° in our setup. The probe spot $\Delta x$ in the sample plane depends on the spectral width $\Delta\omega$ of the probe beam. For a Gaussian pulse, $\Delta x = |q|\Delta\omega f = 2f\sin\alpha/\omega\tau_0$. It should be noted that adding a reflection echelon orthogonal to the grating changes only the probe spot $\Delta y$ in the *y*-direction, so it does not affect the time resolution. Therefore, the extension of the time resolution by the beam cross can be expressed as follows:

$$\Delta t_c = \Delta x \sin\theta / c \qquad (3)$$

In our experiments, the probe spot $\Delta x$ is ~ 0.83 mm and $\theta$ is ~ 33.6°. $\Delta t_c$ can be calculated as 1531fs. In addition, the thickness of the sample can also lead to an extension of the time resolution [2]. Assuming that the sample thickness is *d* and the refraction angle of the pump beam is $\theta_r$, as shown in Fig. S1(b), the extension of the time resolution $\Delta t_s$ due to the sample thickness can be expressed as follows:

$$\Delta t_s = d\left[(\sin\theta_r \sin\theta - 1)/\cos\theta_r + 1\right]/c \qquad (4)$$

Here, $\theta_r$ and $\theta$ are correlated by the law of refraction. In our experiments, the refractive index of ZnSe is ~ 2.7, the thickness *d* is 3 mm, and $\theta_r$ is ~ 11.8°; therefore, $\Delta t_s$ can be calculated as 940 fs. Overall, the equivalent pulse width is $\tau_{probe} = (\tau^2 + \Delta t_c^2 + \Delta t_s^2)^{1/2}$ = 2044 fs. The minimum time resolution $\tau_R$ of a chirped pulse is given by [3]:

$$\tau_R = \sqrt{\tau_0 \sqrt{\tau^2 - \tau_0^2}} \qquad (5)$$

In a long chirp limit, this resolution is equivalent to the minimum time resolution $\tau_R = \sqrt{\tau_0 \tau}$ in a spectrally encoded strategy. Using Eq. (5), we can derive $\tau_R$ = 482fs (811fs, FWHM). We assumed that the pump pulse width $\tau_{pump}$ is maintained at 190 fs (FWHM), that the convolution of the pump pulse and the broadened probe pulse results in a time resolution of $\tau_{con} = (\tau_R^2 + \tau_{pump}^2)^{1/2}$ = 825 fs (FWHM), and that this value is close to the average $\tau_{TPA}$ ~ 855 fs that we measured. Other factors that may broaden the pulse width include the group velocity dispersion (GVD) of the sample, which also depends on the sample thickness. Overall, as seen from the above analysis, to obtain a short time resolution while maintaining a large time window, we need to increase the numerical aperture $NA$ of the converging lens L$_5$ (to reduce the characteristic time $t_c$), reduce the angle $\theta$ between the pump and probe beams, and use a thinner sample.